\documentclass[12pt]{article}
\def\llsymbol#1{\@llsymbol{\@nameuse{c@#1}}}
\def\@llsymbol#1{\ifcase#1\or {}\or {'}\or {''}\or {'''}\or
   {''''}\or {'''''}\or  \else\@ctrerr\fi\relaz}
\newcounter{contador}

\usepackage{color}
\usepackage{indentfirst}
\usepackage{eucal}
\usepackage{amsmath}
\usepackage{amsfonts}
\usepackage{amssymb}
\usepackage{mathrsfs}
\usepackage{bm}
\usepackage{color}
\usepackage{indentfirst}
\usepackage{eucal}
\usepackage{amsmath}
\usepackage{amssymb}
\usepackage{mathrsfs}
\usepackage{bm}
\usepackage{fancyhdr}
\usepackage[nottoc,notlot,notlof]{tocbibind}
\begin{document}
\begin{center}
{\bf Schr\"odinger equation for  two  quasi-exactly solvable 
potentials}\\
%
\vskip 0,4cm {Bartolomeu D. B. Figueiredo \footnote{email:figueiredobartolomeu@gmail.com}}\\
 %
{Centro Brasileiro de Pesquisas F\'{\i}sicas},\\
 {Rua Dr. Xavier Sigaud, 150, CEP 22290-180, Rio de Janeiro, RJ, Brasil}
\end{center}

%
\begin{abstract}
\noindent

\noindent
We apply solutions of Heun's general equation to the 
 stationary Schr\"odinger equation with  two  quasi-exactly solvable 
elliptic potentials which depend on a real parameter $\bm{l}$. 
We get finite-series solutions from power series expansions for Heun's equation if $\bm{l}$ 
is an integer, except if $\bm{l}=-1,-2,-3,-4$.  If  $\bm{l}\neq-5/2$ is half an odd integer, 
we obtain finite series  in terms of hypergeometric 
functions. The quasi-exact
 solvability 
 is expressed by the finite series solutions. However, for any value
of $\bm{l}$,  we find infinite-series eigenfunctions which are convergent and 
bounded for all values of the independent variable.

%
%
\end{abstract}
\tableofcontents
%
%
\section{Introduction }

%




 Extending a previous investigation \cite{eu-2021,eu-2022}, 
 we apply  solutions of  
 Heun's general equation for solving  the one-dimensional 
stationary Schr\"{o}dinger for some potentials.  The 
Schr\"{o}dinger 
equation  is  written as  
\begin{eqnarray}
\label{schr}
\frac{d^2\psi(u)}{du^2}+\big[{\cal E}-{\cal V}(u)\big]\psi(u)=0, 
\end{eqnarray}
where ${\cal E}$ is a constant proportional to the energy of the particle
and ${\cal V}(u)$ is a function obtained from the potential. In the cases 
to be considered,  ${\cal V}(u)$ is given by  
Jacobian elliptic functions and  Eq. 
(\ref{schr}) reduces to instances of the Heun general equation, namely
  \cite{heun,erdelyi-3,maier},
\begin{eqnarray}\label{heun}
\frac{d^{2}H}{dx^{2}}+\left[\frac{\gamma}{x}+
\frac{\delta}{x-1}+\frac{\epsilon}{x-a}\right]\frac{dH}{dx}+\left[
 \frac{\alpha \beta x-q}{x(x-1)(x-a)} \right]H=0, 
 \quad a\in \mathbb{C}\setminus\{0,1\},\vspace{2mm}\\
 \epsilon=\alpha+\beta+1-\gamma-\delta, \nonumber
\end{eqnarray}
 where  
$x=0,1,a,\infty$ are four regular singular points, while $a$, $\alpha$,
$\beta$,  $\gamma$,  $\delta$ and  $q$ are constants.

 Relations among the above equations
 result from the 
substitutions \cite{eu-2021}
\begin{eqnarray}\label{substituicoes}
\begin{array}{l}
a=k^{-2},\qquad x={\rm sn}^2u, \qquad
 H\left[x(u)\right]=\left[{
\rm sn}^2u\right]^{\frac{1-2\gamma}{4}}
 \left[{\rm cn}^2u\right]^{\frac{	1-2\delta}{4}}
\left[ {\rm dn}^2u\right]^{\frac{1-2\epsilon}{4}}Y(u)\end{array}.
\end{eqnarray}
which transform the  Heun equation (\ref{heun}) into the elliptic Darboux equation \cite{darboux,humbert} 
\begin{eqnarray}\label{darboux}
&&\begin{array}{l}
\frac{d^{2}Y(u)}{du^{2}}+\Big\{(\gamma+\delta)^2+1-2\gamma-2\delta-
\left[ 4q-(\gamma+\epsilon)^{2}-1+2\gamma+2\epsilon\right]k^2
- \big[(\alpha-\beta)^{2}
\end{array}
\vspace{2mm}\nonumber\\
&&\begin{array}{l}
-\frac{1}{4}\big] k^2{\rm sn}^2u
- \frac{\left(\gamma-\frac{1}{2} \right)\left( \gamma-\frac{3}{2} \right)}{{\rm sn^2}u}-
 \frac{\left(\delta-\frac{1}{2} \right) \left( \delta-\frac{3}{2} \right){\rm dn^2}u}{{\rm cn^2}u}-
\frac{k^2\left(\epsilon-\frac{1}{2} \right) \left( \epsilon-\frac{3}{2} \right) {\rm cn}^2u}{{\rm dn^2}u}
\Big\}  Y(u)=0,
\end{array}
\end{eqnarray}
where $\operatorname{sn}{u}=\mathrm{sn}(u,k)$,
$\operatorname{cn}{u}=\mathrm{cn}(u,k)$ and $\operatorname{dn}{u}=\operatorname{dn}(u,k)$ are 
the usual Jacobian elliptic functions  modulus $k$ ($0<k^2< 1$) \cite{nist}.
Eq. (\ref{darboux}) becomes an associated Lam\'e equation when 
$\left(\gamma-\frac{1}{2} \right)\left( \gamma-\frac{3}{2} \right)=\left(\delta-\frac{1}{2} 
\right)\left( \delta-\frac{3}{2} \right)=0$, and 
a Lam\'e equation when $\left(\gamma-\frac{1}{2} \right)\left( \gamma-\frac{3}{2} \right)$
$=\left(\delta-\frac{1}{2} \right)\left( \delta-\frac{3}{2} \right)=$$\left(\epsilon-\frac{1}{2} 
\right)\left( \epsilon-\frac{3}{2} \right)=0$.
By ${\rm sn}^2 u+{\rm cn}^{2}u=k^2{\rm sn}^2 u+{\rm dn}^2 u=1$,  Eq. (\ref{schr})
leads to particular cases of (\ref{darboux}) for the following Ganguly's potentials 
\cite{ganguly-1,ganguly-2}:
%
%
\begin{align}
\label{ganguly-1}
&\begin{array}{l}
{\cal V}_1(u)=\frac{2}{\mathrm{sn}^2u}-
\frac{(1-k^2)(\bm{l}+2)(\bm{l}+3)}{\mathrm{dn}^2u},\end{array}\\
\label{ganguly-2}
&\begin{array}{l}
{\cal V}_2(u)=(1-k^2)\left[\frac{2}{\mathrm{cn}^2u}-\frac{(\bm{l}+2)(\bm{l}+3)}
{\mathrm{dn}^2u}\right].\end{array}
\end{align}
%
%
%
Thus, eigenfunctions for the Schr\"{o}dinger 
equation (\ref{schr}) can be obtained systematically  from solutions of the Heun equation 
(\ref{heun}) by means of relations (\ref{substituicoes}) and (\ref{darboux}). Specifically, 
we will find 
 \begin{itemize}
 \itemsep-3pt
\item 
 eigenfunctions in finite series of $x=\operatorname{sn}^2u$  when $\bm{l}$ is 
 integer, excepting $\bm{l}=-1,-2,-3,-4$;
%
\item 
 eigenfunctions in infinite series of $x=\operatorname{sn}^2u$ for any $\bm{l}$;
%
\item 
 eigenfunctions in finite series of hypergeometric functions when $\bm{l}$ is 
 half an odd integer, excepting $\bm{l}=-5/2$.
 \end{itemize}

Problems which admit finite-series solutions
whose coefficients satisfy  three-term or higher order
recurrence relations 
are called  quasi-exactly solvable (QES)
\cite{turbiner1,ushveridze1,kalnins}  
and a part of the energy spectrum, along with the respective 
eigensolutions, can be computed explicitly out of the finite series.
So, we will find that the potentials 
(\ref{ganguly-1}) and (\ref{ganguly-2}) are QES even when $\bm{l}$
is half an odd integer. Furthermore, the presence of finite series
occurs only for particular values of $\bm{l}$ and does
not exclude  infinite-series solutions for any value of $\bm{l}$.

We have already applied solutions of  Heun's equation \cite{eu-2021} 
for a potential ${\cal V}_3(u)$ 
with two parameter $\bm{l}$ and $\bm{m}$ \cite{ganguly-1,ganguly-2,khare,khare2},
%
                 %
%
\begin{eqnarray}\label{v-3}
{\cal V}_3(u)=\bm{m}(\bm{m}+1)\;k^2 \mathrm{sn}^2u+\bm{l}(\bm{l}+1)
 \frac{k^2\;{\rm cn}^2u}{{\rm dn^2}u},
\end{eqnarray}  
for which the Schr\"odinger equation gives an associated Lam\'e equation.              
In fact, to get  eigenfunctions for the potentials  (\ref{ganguly-1}) and  (\ref{ganguly-2}), 
we use some of the 
solutions of Heun's equation that have been constructed for treating the
potential (\ref{v-3}).

 In Section 2 we explain how solutions for the Heun  equation are generated by employing
 transformations of variables which preserve the form of equation (\ref{heun}), and 
 remember some properties of the solutions. In Sections 3 and 4, we discuss  the solutions
 of Schr\"odinger for the potentials  (\ref{ganguly-1}) and  (\ref{ganguly-2}), respectively.
 Concluding remarks are in Section 5. In Appendix A we write the socalled homotopic transformations
 of the Heun equation, while in Appendix B we use such transformations to generate a group of solutions
 which have been omitted in Ref. \cite{eu-2021}.

%



\section{Preliminary Remarks} 
This section summarizes some information on solutions for the Heun
equation (in series of $x$ and of hypergeometric functions) useful for Sections
3 and 4. 
For details see references \cite{eu-2021,eu-2022}.
%

%

%
%

The hypergeometric function
$F(\mathrm{a,b;c};z)$,  defined 
by the series  \cite{erdelyi1}
\begin{eqnarray}\label{hipergeometric}
F\left(\mathrm{a,b;c};z\right)=F(\mathrm{b,a;c};z)&=&1+
\frac{\mathrm{ab}}{1!\ \mathrm{c}}z+
\frac{\mathrm{a(a+1)b(b+1)}}{2!\ \mathrm{c(c+1)}}z^2+\cdots,
\end{eqnarray}
is solution  of the Gauss hypergeometric equation 
\begin{eqnarray*}
\label{hypergeometric}
(1-z)\frac{d^{2}u(z)}{dz^{2}}+\big[\mathrm{c}-(\mathrm{a}+\mathrm{b}+1)z
\big]\frac{du(z)}{dz}-\mathrm{a}\mathrm{b}\;u(z)=0.
\end{eqnarray*}
The series is not defined if the parameter $\mathrm{c}$ is zero or negative integer. 
Furthermore, 
%
\begin{eqnarray}\label{conver-hyper}
F\left(\mathrm{a,b;c};z\right)& & \mbox{converges absolutely for } 
|z|<1,\mbox{ diverges for } |z|>1,\nonumber\\
& &
\mbox{converges also on } |z|=1 \mbox{ if Re(c-a-b)}>0.
\end{eqnarray}
This is valid for infinite series, not for the cases when  
$F\left(\mathrm{a,b;c};z\right)$
reduces to  polynomials \cite{erdelyi1}. We shall also use the Euler relation
\begin{eqnarray}\label{euler-1}
\begin{array}{l}
F(\mathrm{a,b;c};z)=(1-z)^{\mathrm{c-a-b}}
F({c-a,c-b;c;z}), \qquad |z|<1,
\end{array}
\end{eqnarray}
%
%

On the other side, the substitutions of variables of the Heun equation \cite{maier} 
change also the 
parameters 
%
$\left({a},\;q;\; {\alpha},\;{\beta},\;{\gamma},\;{\delta} \right)$ 
of the  equation. 
There are 8 homotopic or elementary-power transformations  which change the dependent variable $H(x)$ 
but  do not modify $x$. In addition, there are 24 fractional linear transformations which map $x$ onto 
$z(x)=(Ax+B)/(Cx+D)$; sometimes, these require a transformation of the dependent variable 
in order to preserve the form of the Heun equation. 

Composition of homotopic and fractional transformations  generate the 192 substitutions of variables
 (including the identity) written in the 
table 2 of Maier's paper  \cite{maier}. 
By enumerating the transformations,  we indicate them by  the symbol $M_i$ ($i=1,2,\cdots,192$)  
which is interpreted as an operator acting on a solution
$H(x)=H(q,a;\alpha,\beta,\gamma,\delta;x)$. 
For example, the fractional substitutions  
\begin{eqnarray}\label{14}
\begin{array}{l}
\frac{x}{a},\qquad
 \frac{(1-a)x}{x-a},
\qquad 1-x, \qquad \frac{a(x-1)}{x-a},\qquad  \frac{a-x}{a},
\end{array}
\end{eqnarray}
correspond to $M_{9}$, $M_{17}$, 
$M_{49}$, $M_{65}$ and $M_{101}$. %
The transformations (\ref{14}) give suitable  arguments $z$
for the
hypergeometric functions $F(\mathrm{a,b;c};z)$ for the Darboux 
equation ($x=\mathrm{sn}^2u$ and $a=1/k^2$) 
because in this case $z$ becomes
\begin{eqnarray*}
\begin{array}{l}
x=\operatorname{sn}^2u,\quad\frac{x}{a}=k^2\operatorname{sn}^2u, 
\quad 1-x=\operatorname{cn}^2u,\vspace{2mm}\\
\frac{(a-1)x}{a-x}=(1-k^2)\frac{\operatorname{sn}^2u}{\operatorname{dn}^2u},
\quad\frac{a(x-1)}{x-a}=\frac{\mathrm{cn}^2u}{\mathrm{dn}^2u},
\quad  \frac{a-x}{a}=\operatorname{dn}^2u
\end{array}
\end{eqnarray*}
and, consequently,  $0\leq z\leq 1$ \cite{eu-2021}. Here we shall need only
the fractional transformation $M_{49}$ which operates as
\begin{align} 
&\begin{array}{l}
M_{49}H(x)=H (1-a,-q +\alpha\beta; \alpha, \beta, \delta, \gamma; 1-x).\quad
\end{array}\label{49}
\end{align}
Besides this,  we shall use
the homotopic transformations, given by $M_{1}-M_{4}$
and ${M}_{25}-M_{28}$ of Maier's table. In Appendix A   
these transformations are denoted by $T_{i}$ ($i=1,\cdots,8$).



We will deal with one group $\mathring{H}^{(i)}(x)$ ($i=1,2,\cdots 8$) of solutions in series of $x$, 
and two groups of solutions
in series of hypergeometric functions, benoted by $\bar{H}^{(i)}(x)$ and $\bm{H}^{(i)}(x)$ 
as in  \cite{eu-2021}.  
The series coefficients 
$b_{n}$ satisfy three-term recurrence relations having
the form
\begin{eqnarray}\label{tres}
\alpha_{0}b_{1}+\beta_{o}b_{0}=0,\qquad
\alpha_{n}b_{n+1}+\beta_{n}b_{n}+\gamma_{n}b_{n-1}=0.
\end{eqnarray}
This system of equations leads to a characteristic equation given by a continued fraction
or by the vanishing of the determinant of a tridiagonal matrix, equation which is satisfied only if there is 
some parameter to be computed, as the energy ${\cal E}$ in Eq. (\ref{schr}).

The solutions $\mathring{H}^{(i)}$ in series of $x$  are given by
%
%
\begin{eqnarray}
 \mathring{H}^{(1)}(x)=\displaystyle \sum_{n=0}^{\infty}
 \mathring{b}_{n}^{(1)}
x^{n},\qquad  \mathring{H}^{(i)}(x)=T_{i} \mathring{H}^{(1)}(x),
\qquad[i=2,3,\cdots,8]
\end{eqnarray}
where the  $\mathring{b}_{n}^{(1)}$ satisfy  the relations
 $\mathring\alpha_{n}^{(1)}\mathring{b}_{n+1}^{(1)}+
 \mathring\beta_{n}^{(1)}\mathring{b}_{n}^{(1)}+
 \mathring\gamma_{n}^{(1)}\mathring{b}_{n-1}^{(1)}=0$ ($\mathring{b}_{-1}^{(1)}=0$)
 with
\begin{eqnarray*}
&& \mathring{\alpha}_n^{(1)}=a(n+\gamma)(n+1),\quad 
  \mathring{\beta}_n^{(1)}=
 -(a+1)n^2-[a(\gamma+\delta-1)+
\alpha+\beta-\delta)]n-q,\nonumber\vspace{2mm}\\
 &&\mathring{\gamma}_n^{(1)}=(n+\alpha-1)(n+\beta-1).
\end{eqnarray*}
The eight expansions of this group are written down in Section 3.1 of
\cite{eu-2021}. In case of infinite series the convergence, according to 
D'Alembetert's ratio test, is given by
%
%
%
\begin{eqnarray}\label{convergence-primeiro}
%
&&\mbox{if } |a|<1,\;\mbox{the series in}\; \mathring{H}^{(i)} \mbox{ converges
	for } |x|<1\;(\Rightarrow x=0 \mbox{ and } x=a); \vspace{3mm}\\ 
&&\mbox{if } |a|>1, \;\mbox{the series in}\;  \mathring{H}^{(i)} \mbox{ converges
	for } |x|<|a|\;(\Rightarrow x=0 \mbox{ and } x=1). 
\end{eqnarray}
In each $\mathring{H}^{(i)}$,  there are multiplicative factors which are important to find 
the behaviour of the expansion at $x=0$, $x=1$ or $x=a$.

To obtain the convergence of infinite series  we have used the ratio  $b_{n+1}/b_n$. 
We find, for example, that for large $n$ \cite{eu-2021},
\begin{eqnarray} 
\begin{array}{l}
  \frac{\mathring{b}_{n+1}^{(1)}}{\mathring{b}_n^{(1)}}\sim\frac{1}{a}\left[
1+\frac{\epsilon-2}{n}\right]\;\quad 
 \text{or }\qquad
  \frac{\mathring{b}_{n+1}^{(1)}}{\mathring{b}_n^{(1)}}\sim 
1+\frac{\delta-2}{n}\;.
\end{array}
\end{eqnarray}
However, if relations (\ref{tres}) satisfy
	\begin{eqnarray*}\label{poinkare-peron}
	\lim_{n\to\infty}\frac{b_{n+1}}{b_n}=t_{1}, \qquad 
	\lim_{n\to\infty}\frac{b_{n+1}}{b_n}=t_{2},\qquad t_1\neq t_2,
	\end{eqnarray*}
	by a Poincar\'e-Perron theorem \cite{gautschi} we have to choose the 
solution with smaller modulus, $|t_1|$ or 
$|t_2|$, in order to assure the convergence of the characteristic equation
(Pincherle theorem). Sometimes this solution is called minimal or recessive \cite{gautschi}. Thus, for minimal solution,  
the convergence for $\mathring{H}^{(1)}$ is given by 
\begin{eqnarray*}\label{conve1}
 \begin{array}{l}\displaystyle
\lim_{n\rightarrow\infty}\;
\vline\frac{\mathring{b}_{n+1}^{(1)}x^{n+1}}{\mathring{b}_{n}^{(1)}x^{n}}\vline=
\begin{cases}
|x|<1,\quad &\mbox{if}\ |a|<1,
\vspace{2mm}
\\
\big| \frac{x}{a}\big|<1, &\mbox{if} \ |a|>1.
\end{cases}
\end{array}
\end{eqnarray*}
This is valid for the other $\mathring{H}^{(i)}$ because  $x$ and $a$ are kept fixed by 
the homotopic transformations $T_i$.
%
%
%

Notice also that \cite{eu-2021},
\begin{eqnarray}\label{ince-dependencia}
\mbox{for infinite series, } \mathring{H}^{(i+4)}(x)=
\mathring{H}^{(i)}(x)\quad\text{if}\quad |x|<|a|\quad \mbox{and}\quad 1-\epsilon\neq 0,1,2,\cdots
\end{eqnarray}
($i=1,2,3,4$). These relations become important in Sections 3.2 and 4.2.
%
%

However, we can have finite series because the series terminates 
on the right-hand side when $\gamma_n=0$ for some $n\geq 1$ 
\cite{arscott} and, then, we have
\begin{eqnarray}\label{truncation}
\begin{array}{l}
 \text{finite series with } 0\leq n\leq N \text{ if }\gamma_{N+1}=0.
\end{array}
\end{eqnarray}
For finite series, if $\beta_i=\mathcal{B}_i-\Lambda$ ($i=0,1,\cdots,N$) 
and if $\alpha_i$, $\mathcal{B}_i$
and $\gamma_i$ are real and independent of $\Lambda$, then 
according to Arscott \cite{arscott,arscott-2}
\begin{eqnarray}\label{autovalores}
\mbox{ there are $N+1$ real 
and distinct values for }\Lambda \;
\mbox{ if }\;\alpha_{i-1}\;\gamma_i>0\;\mbox{ for } i=1,\cdots,N
\end{eqnarray} 
(if $\gamma_i\;\alpha_{i-1}\leq0$
nothing can be said about $\Lambda$).  The values for
$\Lambda$ come from the characteristic equation (see Section 3).

%
%
For $\bar{H}^{(i)}$ and $\bm{H}^{(i)}$ in series of hypegeometric 
functions, we define  $\tilde{F}\left(\mathrm{a,b;c};z\right)$ as
\begin{eqnarray}
\tilde{F}\left(\mathrm{a,b;c};z\right)={F}\left(\mathrm{a,b;c};z\right)/{\Gamma(c)}.
\end{eqnarray}
The expansions $\bar{H}^{(i)}(x)$ are in Appendix B and have the form
\begin{eqnarray}
\label{hiper-zero-1}
\bar{H}^{(1)}(x)= \displaystyle \sum_{n=0}^{\infty}\bar{b}_{n}^{(i)}x^{n}
\tilde{F}\left(n+\alpha,\gamma+\delta-\alpha-1;n+\gamma;
x\right),\qquad \bar{H}^{(i)}(x)=T_i\bar{H}^{(1)}(x), 
\end{eqnarray}
where the argument of the hypergeometric functions is $x$. From $\bar{H}^{(1)}(x)$, we get 
 $\bm{H}^{(i)}(x)$   by the transformations
\begin{eqnarray}
\bm{H}^{(1)}(x)= M_{49}\bar{H}^{(1)}(x),\qquad \bm{H}^{(i)}(x)=T_i\bm{H}^{(1)}(x),
\end{eqnarray}
where the argument of 
the hypergeometric functions is $1-x$. Theses expansions are written in Section 4.3 of Ref. \cite{eu-2021}. 
$\bar{H}^{(i)}$ and $\bm{H}^{(i)}$  will  be used only to get finite
series solutions which must be bounded and convergent for
$0\leq x\leq 1$. By (\ref{conver-hyper}) we see that
\begin{eqnarray}
\begin{array}{l}
\mbox{the hypergeometric functions in }
\bar{{H}}^{(i)}  \; \text{ converge on }
|x|=1 \mbox{ if } \mbox{Re}\;\delta<1; \end{array}
\label{adendos-1-conv}\vspace{2mm}\\
\begin{array}{l}\mbox{hypergeometric } 
\mbox{functions in }
{\bm{H}}^{(i)} \text{ converge on }|x-1|=1  \mbox{ if }
\mbox{Re}\;\gamma<1.\end{array}\label{adendos-2-conv}
\end{eqnarray}
%

%
%
%

%
\section{Solutions for the first Ganguly potential}

The Schr\"{o}dinger
equation (\ref{schr}) for the potencial ${\cal V}_1(u)$, Eq. (\ref{ganguly-1}),  reads
\begin{eqnarray*}
\begin{array}{l}
\frac{d^{2}\psi(u)}{du^{2}}+	
\left[
{\cal E}+2k^2+(\bm{l}+2)(\bm{l}+3)-2\;\frac{\mathrm{dn}^2u}{\mathrm{cn}^2u}-
(\bm{l}+2))(\bm{l}+3)k^2\;\frac{\mathrm{cn}^2u}
{\mathrm{dn}^2u}\right]\psi(u)=0.
\end{array}
\end{eqnarray*}
%
%
This is the Darboux equation (\ref{darboux}) with parameters ($a=1/{k^2}$)
\begin{eqnarray}\label{parametros-ganguly-1}
\begin{array}{l}
 \alpha=\frac{\bm{l}+6}{2}, \quad
\beta=\frac{\bm{l}+5}{2}, \quad \gamma=\frac{1}{2}, \quad
\delta=\frac{5}{2},\quad
q=\frac{\bm{l}^2+6\bm{l}+7}{4}-\frac{\bm{l}^2+5\bm{l}+2+{\cal E}}{4k^2},
\quad \left(\epsilon=\bm{l}+\frac{7}{2}\right).
\end{array}
\end{eqnarray}
Then, by taking $\psi(u)=Y(u)$ in  (\ref{substituicoes}), we get
\begin{eqnarray}\label{psi-ganguly-1}
\psi(u)\stackrel{\text{(\ref{substituicoes})}}{=}
\mathrm{cn}^2 u\;
\mathrm{dn}^{\bm{l}+3} u\;
H[x(u)],\qquad 0 \leq  x(u)={\rm sn}^2 u\leq 1,
\end{eqnarray}
where $H(x)$ is solution of the Heun equation
with the above  parameters. For this potential the appropriate solutions $\psi(u)$ result from the 
expansions $\mathring{H}^{(i)}$ and $\bm{H}^{(i)}$ given, 
respectively, in Sections 3.1 and 4.3 of \cite{eu-2021}.

Eq. (\ref{ganguly-1}) remains invariant under the substitutions  
\begin{eqnarray}\label{simetria-ganguly}
\bm{l}\mapsto -\bm{l}-5.
\end{eqnarray}
This  implies that a same
eigenfunction corresponds to two different values of $\bm{l}$, as we will see.


On the right-hand side, according to (\ref{truncation}), if $\gamma_{n=N+1}=0$, we have
finite series with $0\leq n\leq N$. In the case the recurrence relations take the matrix form
%
\begin{eqnarray}
\label{matriz}
 \left(
\begin{array}{ccccccccc}
\beta_{0} & \alpha_{0} &0   &  &                   \\
\gamma_{1}&\beta_{1}   & \alpha_{1} &                \\
   &  \ddots       &     \ddots        &   \ddots             &          \\
                     & & \gamma_{N-1}& \beta_{N-1}&\alpha_{N-1}\\
                   &   &    0    &\gamma_{N} & \beta_{N}
\end{array}
\right)\left(\begin{array}{l}
b_{0}  \\
b_{1} \\
 \vdots   \\
b_{N-1}\\
b_{N}
\end{array}
\right)=0.
\end{eqnarray}
This system has nontrivial solutions only if the determinant of the
above tridiagonal matrix vanishes. Then, under the Arscott condition
 $\alpha_{i}\gamma_{i+1}>0$, Eq. (\ref{autovalores}), we obtain $N+1$ real 
 and distinct values for the parameter $\Lambda$ (in the present case
 $ \Lambda$ is proportional to ${\cal E}$). For each value there is a solution for $b_n$.

 If $\bm{l}$ is an integer, finite 
series come from  power-series expansions $\mathring{H}^{(i)}$. However, only infinite-series solutions are possible for 
$\bm{l}=-1,-2,-3,-4$ .  If $\bm{l}$ is half an odd integer,  finite series come from expansions $\bm{H}^{(i)}$ 
in series of hypergeometric functions, except 
$\bm{l}=-{5}/{2}$ which corresponds to infinite series.  By requiring  solutions 
bounded  for aIl values of the independent variable $u$, we exclude one
half of the expansions $\mathring{H}^{(i)}$ and $\bm{H}^{(i)}$. In summary, 
we find that
%
%
\begin{itemize}
 \itemsep-3pt
\item  the expansions $\mathring{H}^{(5)}$ and  
$\mathring{H}^{(6)}$ 
in powers of
$x=\operatorname{sn}^2u$ 
lead to finite-series eigenfunctions $\mathring{\psi}_{\bm{l}}^{(5)}$ and  
$\mathring{\psi}_{\bm{l}}^{(6)}$ if $\bm{l}$ s a non-negative integer;
if $\bm{l}$ is an negative integer
less than  $-4$,  
$\mathring{H}^{(1)}$ and  
$\mathring{H}^{(2)}$ 
lead to $\mathring{\psi}_{\bm{l}}^{(1)}$ and  
$\mathring{\psi}_{\bm{l}}^{(2)}$ but these are not independent of $\mathring{\psi}_{\bm{l}}^{(5)}$ and  
$\mathring{\psi}_{\bm{l}}^{(6)}$;
\item  the expansions $\mathring{H}^{(1)}$, $\mathring{H}^{(2)}$, $\mathring{H}^{(5)}$  and  
$\mathring{H}^{(6)}$  also 
lead to infinite-series eigenfunctions denoted by  $\mathring{\Phi}^{(1)}$, 
 $\mathring{\Phi}^{2)}$, $\mathring{\Phi}^{(5)}$  and  
$\mathring{\Phi}^{(6)}$, respectively;
\item  the expansions $\bm{H}^{(5)}$ and  $\bm{H}^{(6)}$ in series of hypergeometric functions lead 
to finite-series eigenfunctions $\bm{\psi}^{(5)}$ and $\bm{\psi}^{(6)}$ if $\bm{l}$ if 
is half an odd integer  greater than or equal to $-3/2$; if 
$\bm{l}$  is half an odd integer  less  than $-5/2$, $\bm{H}^{(1)}$ and  
$\bm{H}^{(2)}$ give $\bm{\psi}^{(1)}$ and $\bm{\psi}^{(2)}$ which are linearly dependent of
 $\bm{\psi}^{(5)}$ and $\bm{\psi}^{(6)}$.
\end{itemize}

We find that the two expansions in finite-series  of
$\operatorname{sn}^2 u$ satisfy Arscott condition 
$\alpha_{n-1} \gamma_n>0$ given in (\ref{autovalores}), but only one expansion in series of 
hypergeometric functions satisfies the condition.

The following properties of the elliptic functions  are useful \cite{nist}:
\begin{align}
&\begin{array}{l}
\mathrm{sn}(-u)=-\mathrm{sn}{\;u},\qquad\qquad
\mathrm{cn}(-u)=\mathrm{cn}{\;u},\qquad\qquad\;
\mathrm{dn}(-u)=\mathrm{dn}{\;u}.
\end{array}
\vspace{2mm}\\
&\begin{array}{l}
\mathrm{sn}(u+2K)=-\mathrm{sn}{\;u},\qquad
\mathrm{cn}(u+2K)=-\mathrm{cn}{\;u},\qquad
\mathrm{dn}(u+2K)=\mathrm{dn}{\;u},
\end{array}
\end{align}
where the period $K$ is given by 
\begin{eqnarray}\label{cd-sd}
K=K(k)=\int_0^{\frac{\pi}{2}}\left[1-k^2 \sin^2 
\theta\right]^{-\frac{1}{2}}d\theta.\nonumber
\end{eqnarray}
We also use the special values 
\begin{eqnarray}
\begin{array}{l}
\mathrm{sn}\;0=0,\quad \mathrm{sn}\;K=1,\quad
\mathrm{cn}\;0=1,\quad \mathrm{cn}{\;K}=0,\quad
\mathrm{dn}\;0=1,\quad \mathrm{dn}{\;K}=1-k^2.
\end{array}
\end{eqnarray}

\subsection{Finite series of $\operatorname{sn}^{2}u$ if $\bm{l}$ is an integer ($\neq -1,-2,-3,-4$)}

First we write the finite-series solutions valid when $\bm{l}$ is a non-negative integer. Inserting the 
expansions $\mathring{H}^{(5)}$ and  
$\mathring{H}^{(6)}$ into (\ref{psi-ganguly-1}) we find even and odd expansions, respectively,
%
\begin{eqnarray}
\begin{cases}
\mathring{\psi}_{\;\bm{l}}^{(5)}(u)
=\displaystyle\;
\frac{\mathrm{cn}^2 u}{\mathrm{dn}^{\bm{l}+2}{u}}
 \sum_{n}
\mathring{b}_n^{(5)}\;\mathrm{sn}^{2n}{u},
\quad 
\begin{cases}
 0\leq n\leq \frac{\bm{l}}{2}\;\; 
\mbox{if}\;\;\bm{l}=0,2,4,\cdots,
\vspace{2mm}\\
0\leq n\leq \frac{\bm{l}-1}{2} \;\;\mbox{if}\;\; \bm{l}=1,3,5,\cdots,
\end{cases} 
\vspace{3mm}\\
\begin{array}{l}
\mathring{\alpha}_n^{(5)}=\frac{1}{
k^2}\left(n+\frac{1}{2}\right)(n+1),\quad \mathring{\beta}_n^{(5)}= -\left(1+\frac{1}{k^2}\right)n^2+
\left(2+\bm{l}-\frac{2}{k^2}\right)n
\vspace{2mm}\end{array}\\
-\frac{{\bm{l}}^2+4\bm{l}+2}{4}+\frac{{\bm{l}}^2+5\bm{l}+2+{\cal E}}{4k^2},\qquad\begin{array}{l}
\mathring{\gamma}_n^{(5)}=\left(n-\frac{\bm{l}+2}{2}\right)\left(n-\frac{\bm{l}+1}{2}\right);
\end{array}
\end{cases}
\end{eqnarray}
%
%
%
\begin{eqnarray}
\begin{cases}
\mathring{\psi}_{\;\bm{l}}^{(6)}(u)
=\displaystyle
\frac{\mathrm{sn}{u}\;\mathrm{cn}^2 u}{
\mathrm{dn}^{\bm{l}+2}{u}}
 \sum_{n}
\mathring{b}_n^{(6)}\;\mathrm{sn}^{2n}{u},
\quad
\begin{cases} 
0\leq n\leq \frac{\bm{l}-1}{2}\;\;
\mbox{if}\;\;  \bm{l}=1,3,5,\cdots,
\vspace{2mm}\\
0\leq n\leq \frac{\bm{l}-2}{2}\;\;  \mbox{if}\;\;  \bm{l}=2,4,6,\cdots,\end{cases}\vspace{3mm}\\
\begin{array}{l}
\mathring{\alpha}_n^{(6)}=\frac{1}{
k^2}\left(n+\frac{3}{2}\right)(n+1),\qquad \mathring{\beta}_n^{(6)}= -\left(1+\frac{1}{k^2}\right)n^2+
\left(1+\bm{l}-\frac{3}{k^2}\right)n
\vspace{2mm}\end{array}\\
-\frac{{\bm{l}}^2+2\bm{l}-1}{4}+\frac{{\bm{l}}^2+5\bm{l}-3+{\cal E}}{4k^2},\qquad\begin{array}{l}
\mathring{\gamma}_n^{(6)}=\left(n-\frac{\bm{l}+1}{2}\right)\left(n-\frac{\bm{l}}{2}\right).
\end{array}
\end{cases}
\end{eqnarray}
These series satisfy the Arscott condition  $\alpha_{n-1}\gamma_{n}> 0$ for $n\geq 1$ 
because for both cases $\alpha_{n-1}>0$ while: 
$\gamma_{n}=\mathring{\gamma}_n^{(5)}\leq 0$
only if $\frac{\bm{l}+1}{2}\leq n\leq\frac{
\bm{l}+2}{2}$, that is, outside the admissible interval for $n$; 
similarly $\gamma_n=\mathring{\gamma}_n^{(6)}\leq 0$
only if $\frac{\bm{l}}{2}\leq n\leq\frac{
\bm{l}+1}{2}$ (outside the admissible interval).


For negative integers $\bm{l}\leq- 5$, putting the  
expansions $\mathring{H}^{(1)}$ and  
$\mathring{H}^{(1)}$ in (\ref{psi-ganguly-1}) we get, respectively,

%
%
%
%
%
\begin{eqnarray}\label{equiv-1}
\begin{array}{ll}
\mathring{\psi}_{\;\bm{l}}^{(1)}(u)
=\displaystyle\;
\mathrm{cn}^2 u\;\mathrm{dn}^{\bm{l}+3}{u}
 \sum_{n}
\mathring{b}_n^{(1)}\;\mathrm{sn}^{2n}{u}=\mathring{\psi}_
{-\bm{l}-5}^{(5)}(u),\;\quad & \bm{l}=-5,-6,-7, \dots;\vspace{2mm}\\
\mathring{\psi}_{\;\bm{l}}^{(2)}(u)=
\displaystyle
{\mathrm{sn}{u}\;\mathrm{cn}^2 u\;\mathrm{dn}^{\bm{l}+3}{u}}
 \sum_{n}
\mathring{b}_n^{(2)}\;\mathrm{sn}^{2n}{u}=\mathring{\psi}_
{-\bm{l}-5}^{(6)}(u), \quad & \bm{l}=-6,-7,-8,\cdots.
\end{array}
\end{eqnarray}
Explicitly, we find  
%
%
\begin{eqnarray*}
\begin{cases}
\mathring{\psi}_{\;\bm{l}}^{(1)}(u)
{=}\displaystyle\;
\mathrm{cn}^2 u\;\mathrm{dn}^{\bm{l}+3}{u}
 \sum_{n}
\mathring{b}_n^{(1)}\;\mathrm{sn}^{2n}{u},
\quad 
\begin{cases}
 0\leq n\leq -\frac{\bm{l}+5}{2}\;\; 
\mbox{if}\;\;\bm{l}=-5,-7,-9,\cdots,
\vspace{2mm}\\
0\leq n\leq -\frac{\bm{l}+6}{2} \;\;\mbox{if}\;\; \bm{l}=-6,-8,-10,\cdots, 
\end{cases} 
\vspace{3mm}\\
\begin{array}{l}
\mathring{\alpha}_n^{(1)}=\frac{1}{
k^2}\left(n+\frac{1}{2}\right)(n+1),\quad \mathring{\beta}_n^{(1)}= -\left(1+\frac{1}{k^2}\right)n^2-
\left(3+\bm{l}+\frac{2}{k^2}\right)n
\vspace{2mm}\end{array}\\
-\frac{{\bm{l}}^2+6\bm{l}+7}{4}+\frac{{\bm{l}}^2+5\bm{l}+2+{\cal E}}{4k^2},\qquad\begin{array}{l}
\mathring{\gamma}_n^{(1)}=\left(n+2+\frac{\bm{l}}{2}\right)\left(n+\frac{\bm{l}+3}{2}\right);
\end{array}
\end{cases}
\end{eqnarray*}
%
%
\begin{eqnarray*}
\begin{cases}
\mathring{\psi}_{\;\bm{l}}^{(2)}(u)
{=}\displaystyle
{\mathrm{sn}{u}\;\mathrm{cn}^2 u\;\mathrm{dn}^{\bm{l}+3}{u}}
 \sum_{n}
\mathring{b}_n^{(2)}\;\mathrm{sn}^{2n}{u},
\quad
\begin{cases} 
0\leq n\leq -\frac{\bm{l}+6}{2}\;\;
\mbox{if}\;\;  \bm{l}=-6,-8,-10,\cdots
\vspace{2mm}\\
0\leq n\leq -\frac{\bm{l}+7}{2}\;\;  \mbox{if}\;\;  \bm{l}=-7,-9,-11,\cdots,\end{cases}\vspace{3mm}\\
\begin{array}{l}
\mathring{\alpha}_n^{(2)}=\frac{1}{
k^2}\left(n+\frac{3}{2}\right)(n+1),\qquad \mathring{\beta}_n^{(2)}= -\left(1+\frac{1}{k^2}\right)n^2-
\left(4+\bm{l}+\frac{3}{k^2}\right)n
\vspace{2mm}\end{array}\\
-\frac{{\bm{l}}^2+8\bm{l}+14}{4}+\frac{{\bm{l}}^2+5\bm{l}-3+{\cal E}}{4k^2},\qquad\begin{array}{l}
\mathring{\gamma}_n^{(2)}=\left(n+\frac{\bm{l}+4}{2}\right)\left(n+\frac{\bm{l+5}}{2}\right).
\end{array}
\end{cases}
\end{eqnarray*}
%
%
%

Eqs. (\ref{equiv-1}) mean that
$\mathring{\psi}_{\;\bm{l}}^{(1)}(u)$ and $\mathring{\psi}_{\;\bm{l}}^{(2)}(u)$  are identical 
to $\mathring{\psi}_{\;\bm{l}}^{(5)}(u)$ and $\mathring{\psi}_{\;\bm{l}}^{(6)}(u)$, respectively:

\begin{eqnarray}\label{equiv-2}
\begin{array}{l}
\mathring{\psi}_{\bm{-5}}^{(1)}(u)
=\mathring{\psi}_
{0}^{(5)}(u),
\qquad\mathring{\psi}_{-6}^{(1)}(u)
=\mathring{\psi}_
{\;1}^{(5)}(u),
\qquad\mathring{\psi}_{-7}^{(1)}(u)
=\mathring{\psi}_
{\;2}^{(5)}(u),
\qquad \dots;\vspace{2mm}\\
\mathring{\psi}_{-6}^{(2)}(u)
=\mathring{\psi}_
{\;1}^{(6)}(u), 
\qquad\mathring{\psi}_{-7}^{(2)}(u)
=\mathring{\psi}_
{\;2}^{(6)}(u), 
\qquad\mathring{\psi}_{-8}^{(2)}(u)
=\mathring{\psi}_
{\;3}^{(6)}(u), \qquad\cdots.
\end{array}
\end{eqnarray}


%
%
%

%
\subsection{Remarks on infinite-series solutions}
The following solutions $\mathring{\Phi}_{\;\bm{l}}^{(i)}(u)$ are given by infinite series:  
%
%
%
%
%
%
\begin{align}
&\mathring{\Phi}_{\;\bm{l}}^{(1)}(u)
=\mathrm{cn}^2 u\;\mathrm{dn}^{{\bm{l}+3}}{u}
\displaystyle\; \;\sum_{n=0}^{\infty}
\mathring{b}_n^{(1)}\;\mathrm{sn}^{2n}{u},
\quad \quad\left[\bm{l}\neq -5,-6,-7,\cdots\right],
\label{inf-1-1}
\vspace{2mm}\\
%
%
%
%
&\mathring{\Phi}_{\;\bm{l}}^{(2)}(u)
{=}\mathrm{sn}{u}\;
\mathrm{cn}^2 u\;\mathrm{dn}^{\bm{l}+3}{u}
\displaystyle\; \;\sum_{n=0}^{\infty}
\mathring{b}_n^{(2)}\;\mathrm{sn}^{2n}{u},\quad [\bm{l}\neq -6,-7,-8;\cdots],\label{inf-2-1}
\vspace{2mm}\\
%
%
%
%
&\mathring{\Phi}_{\;\bm{l}}^{(5)}(u)
{=}\displaystyle\;
\frac{\mathrm{cn}^2 u}{\mathrm{dn}^{\bm{l}+2}{u}}
 \sum_{n=0}^{\infty}
\mathring{b}_n^{(5)}\;\mathrm{sn}^{2n}{u},
\qquad \quad[\bm{l}\neq0,1,2,\cdots],
\label{inf-5-1}\vspace{2mm}\\
%
%
%
%
&\mathring{\Phi}_{\;\bm{l}}^{(6)}(u)
{=}\displaystyle
\frac{\mathrm{sn}{u}\;\mathrm{cn}^2 u}{
\mathrm{dn}^{\bm{l}+2}{u}}
 \sum_{n=0}^{\infty}
\mathring{b}_n^{(6)}\;\mathrm{sn}^{2n}{u},
\qquad\; [\bm{l}\neq1,2,3,\cdots].\label{inf-6-1}
\end{align}
These result from the same expansions used in the previous subsection, but 
now the 
restrictions on $\bm{l}$ assure that $\gamma_n\neq 0$
($n\geq 1$) in each case, as we see from
%
%
\begin{eqnarray*}
&&\begin{array}{l}
\mathring{\gamma}_n^{(1)}=\left(n+2+\frac{\bm{l}}{2}\right)\left(n+1
+\frac{\bm{l}+1}{2}\right),
\end{array}\quad
\begin{array}{l}
\mathring{\gamma}_n^{(2)}=\left(n+2+\frac{\bm{l}+1}{2}\right)\left(n+2
+\frac{\bm{l}}{2}\right),
\end{array}\vspace{2mm}\\
&&
\begin{array}{l}
\mathring{\gamma}_n^{(5)}=\left(n-\frac{\bm{l}+2}{2}\right)\left(n-\frac{\bm{l}+1}{2}\right),
\end{array}
\quad
\begin{array}{l}
\mathring{\gamma}_n^{(6)}=\left(n-\frac{\bm{l}+1}{2}\right)\left(n-\frac{\bm{l}}{2}\right).
\end{array}
\end{eqnarray*}
%


We get the following cases of infinite series:
 \begin{itemize}
 \itemsep-3pt
\item 
 if $\bm{l}=0,1,2,\cdots$,
 expansions $\mathring{\Phi}_{\;\bm{l}}^{(1)}$ and 
 $\mathring{\Phi}_{\;\bm{l}}^{(2)}$ since,  by  (\ref{ince-dependencia}), 
  $\mathring{\Phi}_{\bm{l=0}}^{(6)}=\mathring{\Phi}_{\bm{l=0}}^{(2)}$;
 \item    
 if $\bm{l}=-5,-6,-7,\cdots$,
 expansions $\mathring{\Phi}_{\;\bm{l}}^{(5)}$ and 
 $\mathring{\Phi}_{\;\bm{l}}^{(6)}$ since $\mathring{\Phi}_{\bm{l}=-5}^{(2)}= \mathring{\Phi}_{\bm{l}=-5}^{(6)}$;
 \item    
 if $\bm{l}=-1,-2,-3,-4$ 
 and if 
 $\bm{l}$ is not an integer,   expansions  $\mathring{\Phi}_{\;\bm{l}}^{(1)}$ and 
 $\mathring{\Phi}_{\;\bm{l}}^{(2)}$ which by (\ref{ince-dependencia}) are equivalent 
 to  $\mathring{\Phi}_{\;\bm{l}}^{(5)}$ and 
 $\mathring{\Phi}_{\;\bm{l}}^{(6)}$, respectively. 
 \end{itemize}

Therefore, for each value of $\bm{l}$, there are two solutions 
given by infinite series, regardless of 
the existence or absence of finite-series solutions.


%
%
%
 %
%
%
%

%


\subsection{Finite series of hypergeometric functions if 
$\bm{l}\neq-\frac{5}{2}$ is half an odd integer}
When $\bm{l}$ is half an odd integer 
($\bm{l}\neq -5/2$) we get finite-series eigensolutions bounded for all values of
$x=\operatorname{sn}^2u$  by inserting into (\ref{psi-ganguly-1}) the hypergeometric-series expansions 
${\bm{H}}^{(1)}$, ${\bm{H}}^{(2)}$, ${\bm{H}}^{(5)}$ and ${\bm{H}}^{(6)}$ given in in Section 4.3 of Ref. \cite{eu-2021}.
 From ${\bm{H}}^{(5)}$ and ${\bm{H}}^{(6)}$, 
we find, respectively, 
%

%
%
\begin{eqnarray}\label{adendos-5-associada}
\begin{cases}
\bm{\psi}_{\;\bm{l}}^{(5)}(u)
{=}\displaystyle
\frac{\mathrm{cn}^2u}{ \mathrm{dn}^{\bm{l}+2}u}
\sum_{n=0}^{\bm{l}+\frac{3}{2}}\begin{array}{l} \bm{b}_{n}^{(5)}\mathrm{cn}^{2n}{u}
\;\tilde{F}\left( n-\frac{\bm{l}}{2},2+
\frac{\bm{l}}{2};n+\frac{5}{2};\mathrm{cn}^2{u}\right),\end{array}
\\
\begin{array}{l}\hspace{8cm}\left[ \bm{l}=-\frac{3}{2},-\frac{1}{2},\frac{1}{2},\cdots
\right],
\end{array}
\vspace{3mm}\\
\begin{array}{l}
\bm{\alpha}_{n}^{(5)}=\left(1-\frac{1}{k^2}\right)(n+1),\quad 
\bm{\beta}_{n}^{(5)}=
\left(\frac{1}{k^2}-2\right)n^2+
\left[2+2\bm{l}-\frac{\bm{l}+2}{k^2}\right]n
-\frac{{\cal E}+\bm{l}+2}{4k^2}+
\vspace{3mm}\\
%
%
\frac{2+\bm{l}-\bm{l}^2}{4},\qquad \bm{\gamma}_{n}^{(5)}=
\left(n-\frac{\bm{l}+1}{2}\right)\left(n-\frac{\bm{l}+2}{2}\right)\left(n-\bm{l}-\frac{5}{2}\right),
\end{array}
\end{cases}
\end{eqnarray}
%
%
%
%
%
\begin{eqnarray}\label{adendos-6-associada}
\begin{cases}
\bm{\psi}_{\;\bm{l}}^{(6)}(u)
{=}\displaystyle
\frac{\mathrm{cn}^2u}{ \mathrm{dn}^{\bm{l}+2}u}
\sum_{n=0}^{\bm{l}+3/2}\begin{array}{l} \bm{b}_{n}^{(6)}\mathrm{cn}^{2n}{u}
\;\tilde{F}\left( n+\frac{1-\bm{l}}{2},
\frac{3+\bm{l}}{2};n+\frac{5}{2};\mathrm{cn}^2{u}\right),
\end{array}\\
\begin{array}{l}\hspace{8cm}\left[ \bm{l}=-\frac{3}{2},-\frac{1}{2},\frac{1}{2},\cdots
\right],
\end{array}
\vspace{3mm}\\
\begin{array}{l}
\bm{\alpha}_{n}^{(6)}=\left(1-\frac{1}{k^2}\right)(n+1),\quad 
\bm{\beta}_{n}^{(6)}=
\left(\frac{1}{k^2}-2\right)n^2+
\left[1+2\bm{l}-\frac{\bm{l}+1}{k^2}\right]n
-\frac{{\cal E}+3\bm{l}+5}{4k^2}+
\vspace{3mm}\\
\frac{5+3\bm{l}-\bm{l}^2}{4}, \quad
\bm{\gamma}_{n}^{(6)}=
\left(n-\frac{\bm{l}}{2}\right)\left(n-\frac{\bm{l}+1}{2}\right)\left(n-\bm{l}-\frac{5}{2}\right).
\end{array}
\end{cases}
\end{eqnarray}
where we have used de Euler formula (\ref{euler-1}) to get the same multiplicative 
factor in $\bm{\psi}_{\;\bm{l}}^{(5)}$ and 
$\bm{\psi}_{\;\bm{l}}^{(6)}$. 
In view of (\ref{conver-hyper}), the above hypergeometric functions converge for 
$0\leq \mathrm{cn}^2u\leq1$,  and both expansions are bounded for any $u$. 
For one-term series ($\bm{l}=-3/2$), the solutions are linearly 
dependent
\begin{eqnarray*}
%
%
\bm{\psi}_{\bm{l}=-\frac{3}{2}}^{(5)}(u)\propto
\bm{\psi}_{\bm{l}=-\frac{3}{2}}^{(6)}(u)
{=}\bm{b}_{0}^{(5)}\;
\frac{\mathrm{cn}^2u}{ \sqrt{\mathrm{dn}u}}
\begin{array}{l} 
\;\tilde{F}\left( \frac{3}{4},
\frac{5}{4};\frac{5}{2};\mathrm{cn}^2{u}\right) ,\qquad
{\cal E}=-\frac{1}{2}-\frac{7}{4}k^2.
\end{array}
\end{eqnarray*}
The two-term series ($\bm{l}=-1/2$)  are linearly independent as we see from
\begin{eqnarray*}
\bm{\psi}_{\bm{l}=-\frac{1}{2}}^{(5)}(u)
{=}\bm{b}_{0}^{(5)}
\frac{\mathrm{cn}^2u}{ \mathrm{dn}^{3/2}u}
\begin{array}{l} \left[
\;\tilde{F}\left( \frac{1}{4},
\frac{7}{4};\frac{5}{2};\mathrm{cn}^2{u}\right) -
\frac{\bm\beta_0^{(5)}}{\bm\alpha_0^{(5)}}\mathrm{cn}^{2}{u}
\;\tilde{F}\left( \frac{5}{4},
\frac{7}{4};\frac{7}{2};\mathrm{cn}^2{u}\right) \right]
\end{array}\vspace{2mm}\\
\bm{\psi}_{\bm{l}=-\frac{1}{2}}^{(6)}(u)
{=}\bm{b}_{0}^{(6)}
\frac{\mathrm{cn}^2u}{ \mathrm{dn}^{3/2}u}
\begin{array}{l} \left[
\;\tilde{F}\left( \frac{3}{4},
\frac{5}{4};\frac{5}{2};\mathrm{cn}^2{u}\right) -
\frac{\bm\beta_0^{(6)}}{\bm\alpha_0^{(6)}}\mathrm{cn}^{2}{u}
\;\tilde{F}\left( \frac{5}{4},
\frac{7}{4};\frac{7}{2};\mathrm{cn}^2{u}\right) \right]
\end{array}
\end{eqnarray*}
%
where we have used  the relation ${\alpha_0}b_1=-{\beta_0}b_0$. The previous expansions are 
degenerate because correspond to the same energies since
%
%
\begin{eqnarray*}
\begin{array}{l}
\bm\beta_{0}^{(5)}\bm\beta_{1}^{(5)} 
- \bm\alpha_{0}^{(5)}\bm\gamma_{1}^{(5)}=0,\quad
\bm\beta_{0}^{(6)}\bm\beta_{1}^{(6)} 
- \bm\alpha_{0}^{(6)}\bm\gamma_{1}^{(6)}=0
\;\Rightarrow \vspace{3mm}\\
\hspace{7cm}{\cal E}^{(\pm)}=-\frac{5}{2}-\frac{3}{4}k^2
\pm\sqrt{1+7k^2+k^4}.
\end{array}
\end{eqnarray*}
%
%



Now we regard the condition
$\bm{\alpha}_{n-1}
\bm{\gamma}_{n}>0$.
We have
\begin{eqnarray*}
\begin{array}{ll}
\bm{\alpha}_{n-1}^{(5)}
\bm{\gamma}_{n}^{(5)}=\left(1-\frac{1}{k^2}\right)
\left(n-\bm{l}-\frac{5}{2}\right)n
\left(n-\frac{\bm{l}+1}{2}\right)\left(n-\frac{\bm{l}+2}{2}\right),& 
 \left[1\leq n\leq \bm{l}+\frac{3}{2},\; -\frac{1}{2}, \frac{1}{2},\cdots \right];
 \vspace{2mm}\\
 %
 \bm{\alpha}_{n-1}^{(6)}
\bm{\gamma}_{n}^{(6)}=\left(1-\frac{1}{k^2}\right)
\left(n-\bm{l}-\frac{5}{2}\right)n
\left(n-\frac{\bm{l}}{2}\right)\left(n-\frac{\bm{l}+1}{2}\right),
\;\; &
 \left[1\leq n\leq \bm{l}+\frac{3}{2},\; -\frac{1}{2}, \frac{1}{2}\right].
\end{array}
\end{eqnarray*}
Since $\left(1-\frac{1}{k^2}\right)
\left(n-\bm{l}-\frac{5}{2}\right)n>0$,  $\bm{\alpha}_{n-1}^{(5)}
\bm{\gamma}_{n}^{(5)}\leq 0$ only if  
       $\frac{\bm{l}+1}{2}\leq n\leq\frac{
\bm{l}+1}{2}+ \frac{1}{2}$ for some value of $\bm{l}$,
and $\bm{\alpha}_{n-1}^{(6)}
\bm{\gamma}_{n}^{(6)}\leq 0$ only if  
       $\frac{\bm{l}}{2}\leq n\leq\frac{
\bm{l}}{2}+ \frac{1}{2}$. When these intervals are empty the 
condition is fulfilled.
Thus,  
\begin{eqnarray}
\begin{array}{ll}
\bm{\alpha}_{n-1}^{(5)}
\bm{\gamma}_{n}^{(5)}>0\quad \text{ if } \quad \bm{l}=-\frac{1}{2},\frac{3}{2}, \frac{7}{2}, 
\cdots& \text{ in} \quad\bm{\psi}_{\;\bm{l}}^{(5)}(u),\vspace{3mm}\\
\bm{\alpha}_{n-1}^{(6)}
\bm{\gamma}_{n}^{(6)}>0 \quad\text{ if } \quad
\bm{l}=-\frac{1}{2},\frac{1}{2}, \frac{5}{2}, \frac{9}{2},\cdots 
&\text{ in} \quad \bm{\psi}_{\;\bm{l}}^{(6)}(u).
\end{array}
\end{eqnarray}
Therefore,  for each value value of $\bm{l}$ there is one expansion (two, if $\bm{l}=-1/2$) which satisfies the Arscott condition. 

	
	On the other side,  from ${\bm{H}}^{(1)}$ and ${\bm{H}}^{(2)}$, 
we find $\bm{\psi}_{\;\bm{l}}^{(1)}(u)$ and $\bm{\psi}_{\;\bm{l}}^{(2)}(u)$ for $\bm{l}=-\frac{7}{2},-\frac{9}{2},-\frac{11}{2},\cdots$. 
However these expansions are identical with the previous ones, that is,
%

%
%
\begin{eqnarray*}
&&\bm{\psi}_{\;\bm{l}}^{(1)}(u)
{=}\displaystyle
{\mathrm{cn}^2u}\;
\mathrm{dn}^{\bm{l}+3}u
\sum_{n=0}^{-\bm{l}-7/2}\begin{array}{l} \bm{b}_{n}^{(1)}\mathrm{cn}^{2n}{u}
\;\tilde{F}\left( n+3+\frac{\bm{l}}{2},
-\frac{\bm{l}+2}{2};n+\frac{5}{2};\mathrm{cn}^2{u}\right)
%
\end{array}
\vspace{2mm}\\
&&\begin{array}{l}
\hspace{1.2cm}\stackrel{\text{(\ref{euler-1})}}{=}
\bm{\psi}_{-\bm{l}-5}^{(6)}(u),\qquad\qquad\qquad\left[ \bm{l}=-\frac{7}{2},-\frac{9}{2},-\frac{11}{2},\cdots
\right]\end{array}\nonumber\\
&&\bm{\psi}_{\;\bm{l}}^{(2)}(u)
{=}\displaystyle \operatorname{sn}u\;
{\mathrm{cn}^2u}\;
\mathrm{dn}^{\bm{l}+3}u
\sum_{n=0}^{-\bm{l}-7/2}\begin{array}{l} \bm{b}_{n}^{(2)}\mathrm{cn}^{2n}{u}
\;\tilde{F}\left( n+3+\frac{\bm{l}}{2},
-\frac{\bm{l}}{2};n+\frac{5}{2};\mathrm{cn}^2{u}\right)
\end{array}
\vspace{2mm}\\
&&\begin{array}{l}
\hspace{1.2cm}\stackrel{\text{(\ref{euler-1})}}{=}
\bm{\psi}_{-\bm{l}-5}^{(5)}(u),\qquad\qquad\qquad
\left[ \bm{l}=-\frac{7}{2},-\frac{9}{2},-\frac{11}{2},\cdots
\right]\end{array}.\nonumber
\end{eqnarray*}
%


\section{Solutions for the second Ganguly potential}


The Schr\"{o}dinger
equation (\ref{schr}) for the potencial (\ref{ganguly-2}), Eq (\ref{ganguly-2}), reads
\begin{eqnarray*}
	\begin{array}{l}
		\frac{d^{2}\psi(u)}{du^{2}}+	
		\left[
		{\cal E}+(\bm{l}+2)(\bm{l}+3)-\frac{2}{\mathrm{sn}^2u}-(\bm{l}+2)(\bm{l}+3)k^2\;
		\frac{\mathrm{cn}^2u}
		{\mathrm{dn}^2u}\right]\psi(u)=0,
	\end{array}
\end{eqnarray*}
which  again remains invariant under the substitution  $\bm{l}\mapsto -\bm{l}-5$. 
This is the Darboux equation (\ref{darboux}) with parameters ($a=1/{k^2}$)
\begin{eqnarray}\label{parametros-ganguly-2}
\begin{array}{l}
\alpha=\frac{\bm{l}+6}{2}, \quad
\beta=\frac{\bm{l}+5}{2}, \quad \gamma=\frac{5}{2}, \quad
\delta=\frac{1}{2},\quad 
q=\frac{(\bm{l}+5)^2}{4}-\frac{{\cal E}+2+\bm{l}^2+5\bm{l}}{4k^2},\quad\left( \epsilon=\bm{l}+\frac{7}{2}\right).
\end{array}
\end{eqnarray}
Then, due to (\ref{substituicoes}), $\psi(u)$
follows from solutions of the Heun equation through
\begin{eqnarray}\label{second-ganguly-2}
\qquad\psi[u(x)]\stackrel{\text{(\ref{substituicoes},\ref{parametros-ganguly-2})}}{=}
\mathrm{sn}^2 u\;
\mathrm{dn}^{\bm{l}+3}u\;H(x)].
\end{eqnarray}

This case is analogous to previous one, but now the appropriate solutions come from the 
expansions $\mathring{H}^{(i)}$  given in Section 3.1 of Ref. \cite{eu-2021} 
and from $\bar{H}^{(i)}$ given in Appendix B.  Solutions  
bounded  for all values of the independent variable $u$, are obtained as follows
%
%
\begin{itemize}
 \itemsep-3pt
\item  the expansions $\mathring{H}^{(5)}$ and  
$\mathring{H}^{(7)}$ 
in powers of
$x=\operatorname{sn}^2u$ 
lead to finite-series eigenfunctions $\mathring{\psi}_{\bm{l}}^{(5)}$ and  
$\mathring{\psi}_{\bm{l}}^{(7)}$ if $\bm{l}$ s a non-negative integer;
if $\bm{l}$ is an negative integer
less than  $-4$, 
$\mathring{H}^{(1)}$ and  
$\mathring{H}^{(3)}$  
lead to $\mathring{\psi}_{\bm{l}}^{(1)}$ and  
$\mathring{\psi}_{\bm{l}}^{(3)}$ which are lenearly dependent of  $\mathring{\psi}_{\bm{l}}^{(5)}$ and  
$\mathring{\psi}_{\bm{l}}^{(7)}$.
\item  the expansions $\mathring{H}^{(1)}$, $\mathring{H}^{(3)}$, $\mathring{H}^{(5)}$  and  
$\mathring{H}^{(7)}$  may also 
lead to infinite-series eigenfunctions denoted by  $\mathring{\Phi}^{(1)}$, $\mathring{\Phi}^{3)}$, $\mathring{\Phi}^{(5)}$  and  
$\mathring{\Phi}^{(7)}$, respectively;
\item  the expansions $\bar{H}^{(5)}$ and  $\bar{H}^{(7)}$ in series of hypergeometric functions give
finite-series eigenfunctions $\bar{\psi}^{(5)}$ and $\bar{\psi}^{(7)}$ if $\bm{l}$ if 
is half an odd integer  greater than or equal to $-3/2$; if 
$\bm{l}$  is half an odd integer  less  than $-5/2$, $\bar{H}^{(1)}$ and  
$\bar{H}^{(3)}$ lead to $\bar{\psi}^{(1)}$ and $\bar{\psi}^{(3)}$ which 
are equivalent to $\bar{\psi}^{(5)}$ and $\bar{\psi}^{(7)}$.
\end{itemize}

This time we find that the expansions in series  of
$\operatorname{sn}^2 u$ satisfy Arscott condition $\alpha_{n-1} \gamma_n>0$ which, however,
is not satisfied by expansions in series of hypergeometric functions.

\subsection{Finite series of $\operatorname{sn}^{2}u$ if $\bm{l}$ is an integer ($\neq -1,-2,-3,-4$)}
%


%
From the the solutions $\mathring{H}^{(5)}(x)$ and $\mathring{H}^{(7)}(x)$, we obtain 
two even finite-series expansions for non-negative integers $\bm{l}$,
\begin{eqnarray}
\begin{cases}
\mathring{\psi}_{\;\bm{l}}^{(5)}(u)
{=}\displaystyle\;
\frac{\mathrm{sn}^2 u}{\mathrm{dn}^{\bm{l}+2}{u}}
 \sum_{n}
\mathring{b}_n^{(5)}\;\mathrm{sn}^{2n}{u},
\qquad 
\begin{cases}
 0\leq n\leq \frac{\bm{l}}{2}\;\; 
\mbox{if}\;\;\bm{l}=0,2,4,\cdots,
\vspace{2mm}\\
0\leq n\leq \frac{\bm{l}-1}{2} \;\;\mbox{if}\;\; \bm{l}=1,3,5,\cdots,
\end{cases} 
\vspace{3mm}\\
\begin{array}{l}
\mathring{\alpha}_n^{(5)}=\frac{1}{
k^2}\left(n+\frac{5}{2}\right)(n+1),\qquad
 \mathring{\beta}_n^{(5)}= -\left(1+\frac{1}{k^2}\right)n^2+
\left(\bm{l}-\frac{2}{k^2}\right)n
\vspace{2mm}\end{array}\\
-\frac{{\bm{l}}^2}{4}+\frac{{\bm{l}}^2+5\bm{l}+2+{\cal E}}{4k^2},\qquad\begin{array}{l}
\mathring{\gamma}_n^{(5)}=\left(n-\frac{\bm{l}+1}{2}\right)\left(n-\frac{\bm{l}+2}{2}\right),
\end{array}
\end{cases}
\end{eqnarray}
%
%
%
%
\begin{eqnarray}
\begin{cases}
\mathring{\psi}_{\;\bm{l}}^{(7)}(u)
{=}\displaystyle\;
\frac{\operatorname{cn}u\;\mathrm{sn}^2 u}{\mathrm{dn}^{\bm{l}+2}{u}}
 \sum_{n}
\mathring{b}_n^{(7)}\;\mathrm{sn}^{2n}{u},
\qquad 
\begin{cases}
 0\leq n\leq \frac{\bm{l-1}}{2}\;\; 
\mbox{if}\;\;\bm{l}=1,3,5,\cdots,
\vspace{2mm}\\%
0\leq n\leq \frac{\bm{l}-2}{2} \;\;\mbox{if}\;\; \bm{l}=2,4,6,\cdots,
\end{cases} 
\vspace{3mm}\\
\begin{array}{l}
\mathring{\alpha}_n^{(7)}=\frac{1}{
k^2}\left(n+\frac{5}{2}\right)(n+1),\qquad
 \mathring{\beta}_n^{(7)}= -\left(1+\frac{1}{k^2}\right)n^2+
\left(\bm{l}-\frac{2}{k^2}\right)n
\vspace{2mm}\end{array}\\
-\frac{{\bm{l}}^2}{4}+\frac{{\bm{l}}^2+5\bm{l}-3+{\cal E}}{4k^2},\qquad\begin{array}{l}
\mathring{\gamma}_n^{(7)}=\left(n-\frac{\bm{l}}{2}\right)\left(n-\frac{\bm{l}+1}{2}\right).
\end{array}
\end{cases}
\end{eqnarray}
%
%
%
%
As before, the above series satisfy the condition  $\alpha_{n-1}\gamma_{n}> 0$ 
for $n\geq 1$ 
because $\alpha_{n-1}>0$ while: 
$\gamma_{n}=\mathring{\gamma}_n^{(5)}\leq 0$
only if $\frac{\bm{l}+1}{2}\leq n\leq\frac{
\bm{l}+2}{2}$, that is, outside the admissible interval for $n$; 
similarly $\gamma_n=\mathring{\gamma}_n^{(7)}\leq 0$
only if $\frac{\bm{l}}{2}\leq n\leq\frac{
\bm{l}+1}{2}$ (outside the admissible interval). 


%
%
%
From $\mathring{H}^{(1)}(x)$ and $\mathring{H}^{(3)}(x)$, we obtain 
two  finite-series expansions $\mathring{\psi}_{\;\bm{l}}^{(1)}(u)$ and $\mathring{\psi}_{\;\bm{l}}^{(3)}(u)$ for negative integers $\bm{l}\leq-5$ which, however, are the same as 
$\mathring{\psi}_{\;\bm{l}}^{(5)}(u)$ and $\mathring{\psi}_{\;\bm{l}}^{(7)}(u)$, since
\begin{eqnarray}
\begin{array}{ll}
 \mathring{\psi}_{\;\bm{l}}^{(1)}(u)=\displaystyle \mathrm{sn}^2 u\;
\mathrm{dn}^{\bm{l}+3}u\sum_{n}
 \mathring{b}_{n}^{(1)}
\operatorname{sn}^{2n}u=\mathring{\psi}_{-\bm{l}-5}^{(5)}(u),&
\qquad[\bm{l}=-5,-6,-7,\cdots],
\vspace{2mm}\\
%
\mathring{\psi}_{\,\bm{l}}^{(3)}(u)=
\operatorname{cn}u\;\mathrm{sn}^2 u\;
\mathrm{dn}^{\bm{l}+3}u\displaystyle \sum_{n}\mathring{b}_{n}^{(3)}
\operatorname{sn}^{2n}u=\mathring{\psi}_{-\bm{l}-5}^{(7)}(u),
&\qquad[\bm{l}=-6,-7,-8,\cdots],
\end{array}
\end{eqnarray}
where $\mathring{\gamma}_n^{(1)}=\left(n+\frac{\bm{l}+3}{2}\right)\left(n+\frac{\bm{l}+4}{2}\right),\;
\mathring{\gamma}_n^{(3)}=\left(n+\frac{\bm{l}+4}{2}\right)\left(n+\frac{\bm{l}+5}{2}\right)$.
%
%
%
%

%
\subsection{Remarks on infinite-series solutions}

   From the power series solutions        
%
%
%
\begin{align}
&\begin{array}{ll}
 \mathring{\Phi}_{\;\bm{l}}^{(1)}(u)=\displaystyle \mathrm{sn}^2 u\;
\mathrm{dn}^{\bm{l}+3}u\sum_{n=0}^{\infty}
 \mathring{b}_{n}^{(1)}
\operatorname{sn}^{2n}u,&
\hspace{0.8cm}[\bm{l}\neq-5,-6,-7,\cdots],\end{array}
\vspace{2mm}\\
&\begin{array}{l}
\mathring{\Phi}_{\,\bm{l}}^{(3)}(u)=
\operatorname{cn}u\;\mathrm{sn}^2 u\;
\mathrm{dn}^{\bm{l}+3}u\displaystyle \sum_{n=0}^{\infty}\mathring{b}_{n}^{(3)}
\operatorname{sn}^{2n}u,
\quad[\bm{l}\neq-6,-7,-8,\cdots],
\end{array}
\end{align}
%
%
%
\begin{align}
&
\mathring{\Phi}_{\;\bm{l}}^{(5)}(u)
{=}\displaystyle\;
\frac{\mathrm{sn}^2 u}{\mathrm{dn}^{\bm{l}+2}{u}}
 \sum_{n=0}^{\infty}
\mathring{b}_n^{(5)}\;\mathrm{sn}^{2n}{u},
\qquad\quad [\bm{l}\neq 0, 1,2,\cdots],\vspace{2mm}\\
&\mathring{\Phi}_{\;\bm{l}}^{(7)}(u)
{=}\displaystyle\;
\frac{\operatorname{cn}u\;\mathrm{sn}^2 u}{\mathrm{dn}^{\bm{l}+2}{u}}
 \sum_{n=0}^{\infty}
\mathring{b}_n^{(7)}\;\mathrm{sn}^{2n}{u},
\qquad [\bm{l}\neq 1,2,3,\cdots].
\end{align}

Thence, we get the following cases of infinite series:
 \begin{itemize}
 \itemsep-3pt
\item 
 if $\bm{l}=0,1,2,\cdots$,
 expansions $\mathring{\Phi}_{\;\bm{l}}^{(1)}$ and 
 $\mathring{\Phi}_{\;\bm{l}}^{(3)}$ since $\mathring{\Phi}_{\bm{l}=0}^{(7)}=\mathring{\Phi}_{\bm{l}=0}^{(3)}$ ;
 \item    
 if $\bm{l}=-5,-6,-7,\cdots$,
 expansions $\mathring{\Phi}_{\;\bm{l}}^{(5)}$ and 
 $\mathring{\Phi}_{\;\bm{l}}^{(7)}$ since $\mathring{\Phi}_{\bm{l}=-5}^{(3)}=
\mathring{\Phi}_{\bm{l}=-5}^{(7)} $;
 \item    
 if $\bm{l}=-1,-2,-3,-4$ 
 and if 
 $\bm{l}$ is not an integer,   expansions  $\mathring{\Phi}_{\;\bm{l}}^{(1)}$ and 
 $\mathring{\Phi}_{\;\bm{l}}^{(3)}$ which by (\ref{ince-dependencia}) are equivalent to  $\mathring{\Phi}_{\;\bm{l}}^{(5)}$ and 
 $\mathring{\Phi}_{\;\bm{l}}^{(7)}$, respectively. 
 \end{itemize}
Thus, for each $\bm{l}$ there are two infinite-series expansions, even for the cases  admitting finite-series  
solutions.

\subsection{Finite series of hypergeometric functions if 
$\bm{l}\neq-\frac{5}{2}$ is half an odd integer}
When $\bm{l}$ is half an odd integer 
($\bm{l}\neq -5/2$) we get finite series  of hypergeometric functions, bounded for all values of
$x=\operatorname{sn}^2u$,  by inserting into (\ref{second-ganguly-2}) the expansions 
${\bar{H}}^{(1)}$, ${\bar{H}}^{(3)}$,  
${\bar{H}}^{(5)}$ and ${\bar{H}}^{(7)}$. From ${\bar{H}}^{(5)}$ and ${\bar{H}}^{(7)}$ we 
find, respectively, 
%
%
%
%
%
\begin{eqnarray}
\begin{cases}
\bar{\psi}_{\;\bm{l}}^{(5)}(u)
\stackrel{\text{(\ref{bar-5}})}{=}
\displaystyle
\frac{\mathrm{sn}^2u}{ \mathrm{dn}^{\bm{l}+2}u}
\sum_{n=0}^{\bm{l}+3/2}\begin{array}{l} \bar{b}_{n}^{(5)}\mathrm{sn}^{2n}{u}
\;\tilde{F}\left( n-\frac{\bm{l}}{2},2+
\frac{\bm{l}}{2};n+\frac{5}{2};\mathrm{sn}^2{u}\right),\end{array}
\vspace{2mm}\\
\begin{array}{l}
\hspace{7.8cm}
\left[ \bm{l}=-\frac{3}{2},-\frac{1}{2},\frac{1}{2},\cdots
\right],\end{array}
\vspace{3mm}\\
\begin{array}{l}
\bar{\alpha}_{n}^{(5)}=\frac{1}{k^2}(n+1),\qquad
\bar{\beta}_{n}^{(5)}=-
\left(1+\frac{1}{k^2}\right)n^2+
\left(\bm{l}+\frac{\bm{l}+2}{k^2}\right)n+\frac{{\cal E}+\bm{l}+2}{4k^2}-\frac{\bm{l}^2}{4}, 
\vspace{3mm}\\
%
%
%
\bar{\gamma}_{n}^{(5)}=
\left(n-\frac{\bm{l}+1}{2}\right)\left(n-\frac{\bm{l}+2}{2}\right)\left(n-\bm{l}-\frac{5}{2}\right),
\end{array}
\end{cases}
\end{eqnarray}
%
%
%
%
\begin{eqnarray}
\begin{cases}
\bar{\psi}_{\;\bm{l}}^{(7)}(u)
\stackrel{\text{(\ref{bar-7}})}{=}
\displaystyle
\frac{\operatorname{cn}u\;\mathrm{sn}^2u}{ \mathrm{dn}^{\bm{l}+2}u}
\sum_{n=0}^{\bm{l}+3/2}\begin{array}{l} \bar{b}_{n}^{(5)}\mathrm{sn}^{2n}{u}
\;\tilde{F}\left( n-\frac{\bm{l}-1}{2},
\frac{\bm{l}+3}{2};n+\frac{5}{2};\mathrm{cn}^2{u}\right)\end{array},
\vspace{2mm}\\
\begin{array}{l}\hspace{7.8cm}
\quad
\left[ \bm{l}=-\frac{3}{2},-\frac{1}{2},\frac{1}{2},\cdots
\right]\end{array}
\vspace{3mm}\\
\begin{array}{l}
\bar{\alpha}_{n}^{(7)}=\frac{1}{k^2}(n+1),\qquad
\bar{\beta}_{n}^{(7)}=-
\left(1+\frac{1}{k^2}\right)n^2+
\left(\bm{l}+\frac{\bm{l}+1}{k^2}\right)
n+\frac{{\cal E}+3\bm{l}+5}{4k^2}-\frac{\bm{l}^2}{4}, 
\vspace{3mm}\\
%
%
%
\bar{\gamma}_{n}^{(7)}=
\left(n-\frac{\bm{l}}{2}\right)\left(n-\frac{\bm{l}+1}{2}\right)\left(n-\bm{l}-\frac{5}{2}\right).
\end{array}
\end{cases}
\end{eqnarray}
We obtain
\begin{eqnarray*}
&&\begin{array}{l}
		\bar\alpha_{n-1}^{(5)}\bar\gamma_n^{(5)}=
		\frac{1}{k^2}{\; n\left(n-\bm{l}-\frac{5}{2}\right) }
		\left(n-\frac{\bm{l}+1}{2}\right)
		\left(n-\frac{\bm{l}+2}{2}\right)
		\quad \mbox{for}\quad 1\leq n\leq \bm{l}+\frac{3}{2},
	\end{array}\vspace{3mm}\\
	&&\begin{array}{l}
		\bar\alpha_{n-1}^{(7)}\bar\gamma_n^{(7)}=
		\frac{1}{k^2}{\; n\left(n-\bm{l}-\frac{5}{2}\right) }
		\left(n-\frac{\bm{l}}{2}\right)
		\left(n-\frac{\bm{l}+1}{2}\right)
		\quad \mbox{for}\quad 1\leq n\leq \bm{l}+\frac{3}{2},
	\end{array}
\end{eqnarray*} 
Then, the Arscott condition is not satisfied because
\begin{eqnarray*}
&&\begin{array}{l}
		\bar\alpha_{n-1}^{(5)}\bar\gamma_n^{(5)}>0\quad\text{only if}\quad
		\frac{\bm{l}+1}{2}<n<
	\frac{\bm{l}+1}{2}+\frac{1}{2}
	\end{array},\vspace{3mm}\\
	&&\begin{array}{l}
		\bar\alpha_{n-1}^{(7)}\bar\gamma_n^{(7)}>0\quad\text{only if}\quad
		\frac{\bm{l}}{2}<n<
	\frac{\bm{l}}{2}+\frac{1}{2}
	\end{array},
\end{eqnarray*} 
which include at most for one value of $n$. Nevertheless,  each one-term ($\bm{l}=-3/2$) and two-term ($\bm{l}=-1/2$) series solutions 
present real energies and are degenerate. We obtain
\begin{eqnarray}
&&\begin{array}{l}
{\bar\beta}_0^{(5)}=0,\;  {\bar\beta}_0^{(7)}=0,\quad\Rightarrow \quad 
{\cal E}=-\frac{1}{2}+ \frac{9}{4}\;k^2\quad\mbox{for} \quad
\bar{\psi}_{\bm{l}=-\frac{3}{2}}^{(5)},\;\bar{\psi}_{\bm{l}=-\frac{3}{2}}^{(7)};\end{array}\vspace{2mm}\\
&&\begin{array}{l}
\bar\beta_{0}^{(5)}\bar\beta_{1}^{(5)} 
- \bar\alpha_{0}^{(5)}\bar\gamma_{1}^{(5)}=0,
\;
\bar\beta_{0}^{(7)}\bar\beta_{1}^{(7)} 
- \bar\alpha_{0}^{(7)}\bar\gamma_{1}^{(7)}=0
\quad\Rightarrow \end{array}\nonumber\vspace{3mm}\\
&&\begin{array}{l}\qquad 
{\cal E}^{(\pm)}=-\frac{5}{2}+\frac{13}{4}k^2
\pm\sqrt{1-9k^2+9k^4} \quad\mbox{for} \quad
\bar{\psi}_{\bm{l}=-\frac{1}{2}}^{(5)},\;\bar{\psi}_{\bm{l}=-\frac{1}{2}}^{(7)}.
\end{array}
\end{eqnarray}

From the expansions ${\bar{H}}^{(5)}$ and ${\bar{H}}^{(7)}$ 
we obtain solutions for $\bm{l}\leq -7/2$, but such solutions reproduce  
$ \bar{\psi}_{\;\bm{l}}^{(7)}$ and $\bar{\psi}_{\;\bm{l}}^{(5)}$.
respectively. Indeed,

%
%
\begin{eqnarray}%
&&\bar{\psi}_{\;\bm{l}}^{(1)}(u)\stackrel{\text{(\ref{bar-1}})}{=} 
\operatorname{sn}^2u \operatorname{dn}^{\bm{l}+3}u\displaystyle \sum_{n=0}^{-\bm{l}-\frac{7}{2}}\bar{b}_{n}^{(1)}\operatorname{sn}^{2n}u
\begin{array}{l}\tilde{F}\left(n+\frac{\bm{l}+6}{2}, -\frac{\bm{l}+2}{2};n+\frac{5}{2};
\operatorname{sn}^2u\right)\nonumber
\end{array}
\vspace{2mm}\\
&&\begin{array}{l}
\hspace{1.2cm}\stackrel{\text{(\ref{euler-1}})}{=}
\bar{\psi}_{-\bm{l}-5}^{(7)}(u),\hspace{3cm} 
\left[\bm{l}=-\frac{7}{2},-\frac{9}{2},-\frac{11}{2},\cdots \right],
\end{array}\vspace{3mm}\\
&&\bar{\psi}_{\;\bm{l}}^{(3)}(u)\stackrel{\text{(\ref{bar-3}})}{=}\operatorname{cu}u\; 
\operatorname{sn}^2u \operatorname{dn}^{\bm{l}+3}u\displaystyle \sum_{n=0}^{-\bm{l}-\frac{7}{2}}\bar{b}_{n}^{(1)}\operatorname{sn}^{2n}u
\begin{array}{l}\tilde{F}\left(n+\frac{\bm{l}+6}{2}, -\frac{\bm{l}}{2};n+\frac{5}{2};
\operatorname{sn}^2u\right)\nonumber
\end{array}
\vspace{2mm}\\
&&\begin{array}{l}
\hspace{1.2cm}\stackrel{\text{(\ref{euler-1}})}{=}
\bar{\psi}_{-\bm{l}-5}^{(5)}(u),
\hspace{3cm} \left[\bm{l}=-\frac{7}{2},-\frac{9}{2},-\frac{11}{2},\cdots \right],
\end{array}
\end{eqnarray}
%

%
	\section{Conclusions}

By using solutions of the Heun general equation (\ref{heun}), we have seen that the  
potentials (\ref{ganguly-1}) and (\ref{ganguly-2}) are quasi-exactly solvable (QES)
 when $\bm{l}$ is 
 integer, excepting $\bm{l}=-1,-2,-3,-4$; however they are QES 
 even when
$\bm{l}\neq -5/2$ is half an odd integer. Furthermore, this procedure also gives two convergent infinite 
series for each value of the parameter $\bm{l}$, a fact which 
does not follow from the previous approaches.

In order to  obtain the  infinite series we have used four
of the eight solutions $\mathring{H}^{(i)}(x)$ (resulting from homotopic transformations) in powers of $x$ for the Heun equation; the remaining 
solutions contain unbounded multiplicative factors. Beside this,  in the present cases,  the range of the independent 
variable $x=\operatorname{sn}^2u$ lies inside the region of convergence 
of $\mathring{H}^{(i)}(x)$ which includes only two singular points, $x=0$ and $x=1$.
 
On the other side, we have already found \cite{eu-2021} that expansions in powers series and in series
of hypergeometric functions for Heun's general equation  lead to solutions for the confluent Heun equation (CHE) (also called generalized spheroidal equation \cite{fisher,leaver}). Then, 
we can ask if there are QES potentials which, in addition to finite-series solutions, also
admit infinite series as the ones aforementioned.

In fact, for two  Ushveridge's  trigonometric potentials \cite{ushveridze1}, 
the Schr\"odinger equation (\ref{schr}) obeys a CHE \cite{eu-lea}. Thence, 
to treat the above problem, first we have to generate new solutions by using the appropriate transformations of the CHE \cite{decarreau1,decarreau2}. Indeed, 
it seems that these transformations will generate  finite-series and infinite-series 
solutions for such QES potentials.

\section*{Acknowledgement}
 
%
This work was funded by Minist\'erio da Ci\^encia, Tecnologia e Inova\c{c}\~ oes do Brasil. The author thanks 
L\'ea Jaccoud El-Jaick for reading the manuscript. This work does not have any conflicts of interest.

\section*{Data availibility} There are no associated data concerning this work. 
%
\appendix
\section{Homotopic transformations}
\protect\label{A}
\setcounter{equation}{0}
\renewcommand{\theequation}{B.\arabic{equation}}
The homotopic transformations, given by $M_{1}-M_{4}$
and ${M}_{25}-M_{28}$ of Maier's table are denoted by 
$T_{i}$ ($i=1,\cdots,8$) according to
\begin{eqnarray*}
\begin{array}{l}
T_1=M_1,\;
T_2 =M_{25}, \;
T_3=M_{2},\;
T_4=M_{26},\;
T_5=M_3,\;
T_6=M_{27},\;
T_7=M_4,\;
T_8 =M_{28}.\end{array}
\end{eqnarray*}
They are given by
\begin{align}
&\begin{array}{l}
T_1H(x)=H(x)=H(a,q;\alpha,\beta,\gamma,\delta;x),\quad (\text{identity})
\end{array}\label{t1}
\vspace{2mm}\\
&\begin{array}{l}
T_{2}H(x)=
x^{1-\gamma}H\big[a,q-(\gamma-1)(\delta a+\epsilon);{\beta-\gamma+1},
\alpha-\gamma+1,2-\gamma,\delta;x \big],
\end{array}\label{t2}
\vspace{2mm}\\
%
&\begin{array}{l}
T_{3}H(x)= (1-x)^{1-\delta}H\big[a,q-(\delta-1)\gamma a;{\beta-\delta+1},
\alpha-\delta+1,\gamma,2-\delta;x \big],\end{array}\label{t3}
\vspace{2mm}\\
%
%
&\begin{array}{l}
T_{4}H(x)= x^{1-\gamma}(1-x)^{1-\delta}
H\big[a,q-(\gamma+\delta-2)a-(\gamma-1)\epsilon;
\vspace{2mm}\\
\hspace{2.0cm}
{\alpha-\gamma-\delta+2},
\beta-\gamma-\delta+2,2-\gamma,2-\delta;x \big],\end{array}
\label{t4}
\end{align}
%
%
\begin{align}
&\begin{array}{l}
T_{5}H(x)=
\left[1-\frac{x}{a} \right]^{1-\epsilon}
H\big[a,q-\gamma(\alpha+\beta-\gamma-\delta);
{-\alpha+\gamma+\delta},
-\beta+\gamma+\delta,
\gamma,\delta;x \big],\end{array}
\label{t5}
\vspace{2mm}\\
&\begin{array}{l}
T_{6}H(x)= \left[1-\frac{x}{a} \right]^{1-\epsilon} x^{1-\gamma}
H\big[a,q-\delta(\gamma-1)a-\alpha-\beta+\delta+1;
\vspace{2mm}\\
\hspace{2cm}
{-\beta+\delta+1},-\alpha+\delta+1,2-\gamma,\delta;x \big],
\end{array}
\label{t6}\vspace{2mm}\\
 &\begin{array}{l}
 T_{7}H(x)=  \left[1-\frac{x}{a} \right]^{1-\epsilon}(1-x)^{1-\delta}
H\big[a,q-\gamma[(\delta-1)a+\alpha+\beta-\gamma-\delta];
\vspace{2mm}\\
\hspace{2cm}
{-\beta+\gamma+1},-\alpha+\gamma+1,\gamma,2-\delta;x \big],
\end{array}\label{t7}
\vspace{2mm}\\
&\begin{array}{l}
T_{8}H(x)=\left[1-\frac{x}{a} \right]^{1-\epsilon}
\displaystyle x^{1-\gamma}
(1-x)^{1-\delta}
H\big[a,q-(\gamma+\delta-2)a
-\alpha-\beta+\delta+1;
\vspace{2mm}\\
\hspace{2cm}
{2-\alpha},2-\beta,2-\gamma,2-\delta;x \big].
\end{array}\label{t8}
\end{align}
%


%
%
%
\section{Expansions $\bar{H}^{(i)}(x)$ in series of hypergeometric functions }
\protect\label{B}
\setcounter{equation}{0}
\renewcommand{\theequation}{B.\arabic{equation}}

The initial solution $\bar{H}^{(1)}(x)$ has been constructed in Ref. \cite{eu-2021}. In the following
we write all the expansions $\bar{H}^{(i)}(x)=T_i\bar{H}^{(1)}(x)$ because they are used in Section 4.

%
\begin{eqnarray}\label{bar-1}
\label{hiper-zero-1}\begin{cases}
\bar{H}^{(1)}(x)= \displaystyle \sum_{n=0}^{\infty}\bar{b}_{n}^{(1)}x^{n}
\tilde{F}\left(n+\alpha,\gamma+\delta-\alpha-1;n+\gamma;
x\right), \vspace{3mm}\\
%
¨
\bar{\alpha}_{n}^{(1)}=a(n+1),\qquad
\bar{\beta}_{n}^{(1)}=-(a+1)n^2-[a(2\alpha+1-\gamma-\delta)+\alpha+\beta-\delta]n
-q \vspace{2mm}\\
-a\alpha (\alpha+1-\gamma-\delta),\qquad
%
\bar{\gamma}_{n}^{(1)}=(n+\alpha-1)(n+\alpha-\delta)(n+\alpha+\beta-\gamma-\delta)
. \end{cases}
\end{eqnarray}
%
%
%
\begin{eqnarray}\label{bar-2}
\label{hiper-zero-2}\begin{cases}
\bar{H}^{(2)}(x)= x^{1-\gamma}\displaystyle \sum_{n=0}^{\infty}
\bar{b}_{n}^{(2)}x^{n}
\tilde{F}\left(n+\beta-\gamma+1,\delta-\beta;n+2-\gamma;
x\right)\quad  \vspace{3mm}\\
%
\bar{\alpha}_{n}^{(2)}=a(n+1),\qquad
\bar{\beta}_{n}^{(2)}=-(a+1)n^2-[a(2\beta-\gamma-\delta)+\alpha+\beta+\vspace{2mm}\\
2-2\gamma-\delta]n
-q -a(\beta+1-\gamma) (\beta-\delta)+(\gamma-1)(\delta a+\epsilon),\vspace{2mm}\\
%
\bar{\gamma}_{n}^{(2)}=(n+\beta-\gamma)(n+\beta+1-\gamma-\delta)(n+\alpha+\beta-\gamma-\delta)
. \end{cases}
\end{eqnarray}
%
%
%
The expansions $\bar{H}^{(3)}$ and $\bar{H}^{(4)}$ can be obtained from the previous ones as
\begin{eqnarray}
\label{bar-3}
\bar{H}^{(3)}(x)&= &(1-x)^{1-\delta}\displaystyle \sum_{n=0}^{\infty}
\bar{b}_{n}^{(3)}x^{n}
\tilde{F}\left(n+\beta-\delta+1,\gamma-\beta;n+\gamma;
x\right)\nonumber\vspace{2mm}\\
&\stackrel{\text{(\ref{euler-1}})}{=}&\bar{{H}}^{(1)}(x)\Big|_{\alpha\leftrightarrow\beta},
\end{eqnarray}
\begin{eqnarray}
\label{bar-4}
\bar{H}^{(4)}(x)&=&x^{1-\gamma} (1-x)^{1-\delta}\displaystyle \sum_{n=0}^{\infty}\bar{b}_{n}^{(4)}x^{n}
\tilde{F}\left(n+\alpha-\gamma-\delta+2,1-\alpha;n+2-\gamma;
x\right)\nonumber\vspace{2mm}\\
&\stackrel{\text{(\ref{euler-1}})}{=}&\bar{{H}}^{(2)}(x)\Big|_{\alpha\leftrightarrow\beta}.
\end{eqnarray}
where $\alpha$ and $\beta$ must be interchanged also in the recurrence relations.

%

%
%
\begin{eqnarray}%
\label{bar-5}\begin{cases}
\bar{H}^{(5)}(x)= \left(1-\frac{x}{a}\right)^{1-\epsilon}\displaystyle \sum_{n=0}^{\infty}\bar{b}_{n}^{(5)}x^{n}
\tilde{F}\left(n-\alpha+\gamma+\delta,\alpha-1;n+\gamma;
x\right), \vspace{3mm}\\
%
¨
\bar{\alpha}_{n}^{(5)}=a(n+1),\qquad
\bar{\beta}_{n}^{(5)}=-(a+1)n^2-[a(1-2\alpha+\gamma+\delta)-\alpha-\beta
\vspace{3mm}\\
%
+2\gamma+\delta]n-q+\gamma(\alpha+\beta-\gamma-\delta)-a(\alpha-1) (\alpha-\gamma-\delta),  \vspace{2mm}\\
\bar{\gamma}_{n}^{(5)}=(n-\alpha-1+\gamma+\delta)
(n-\alpha+\gamma)(n-\alpha-\beta+\gamma+\delta)
. \end{cases}
\end{eqnarray}
%
%
%
\begin{eqnarray}%
\label{bar-6}\begin{cases}
\bar{H}^{(6)}(x)= x^{1-\gamma}\left(1-\frac{x}{a}\right)^{1-\epsilon}\displaystyle \sum_{n=0}^{\infty}\bar{b}_{n}^{(6)}x^{n}
\tilde{F}\left(n-\beta+\delta+1,\beta-\gamma;n+2-\gamma;
x\right), \vspace{3mm}\\
%
¨
\bar{\alpha}_{n}^{(6)}=a(n+1),\qquad
\bar{\beta}_{n}^{(6)}=-(a+1)n^2-[a(1-2\beta+\gamma+\delta)-\alpha-\beta+
\vspace{2mm}\\
\delta-+2]n-q +\alpha+\beta-\delta-1-a(\beta-1)(\beta-\gamma-\delta),\vspace{2mm}\\
%
\bar{\gamma}_{n}^{(6)}=(n-\beta+\delta)(n-\beta+1)(n-\alpha-\beta+\gamma+\delta)
. \end{cases}
\end{eqnarray}
For $\bar{H}^{(7)}$ and $\bar{H}^{(8)}$ we get
%
%
\begin{eqnarray}%
\label{bar-7}
\bar{H}^{(7)}(x)&=&(1-x)^{1-\delta}\left(1-\frac{x}{a}\right)^{1-\epsilon}\displaystyle \sum_{n=0}^{\infty}\bar{b}_{n}^{(7)}x^{n}
\tilde{F}\left(n-\beta+\gamma+1,\beta-\delta;n+\gamma;
x\right)\nonumber\vspace{2mm}\\
&\stackrel{\text{(\ref{euler-1}})}{=}&\bar{{H}}^{(5)}(x)\Big|_{\alpha\leftrightarrow\beta}, 
\end{eqnarray}
%
%
%
%
\begin{eqnarray}
\label{bar-8}
\bar{H}^{(8)}(x)&=& x^{1-\gamma}
(1-x)^{1-\delta}\left(1-\frac{x}{a}\right)^{1-\epsilon}\displaystyle \sum_{n=0}^{\infty}{b}_{n}^{(8)}x^{n}
\qquad\nonumber\\
&\times&
\tilde{F}\left(n+2-\alpha,\alpha-
\gamma-\delta+1;n+2-\gamma;
x\right)\stackrel{\text{(\ref{euler-1}})}{=}\bar{{H}}^{(6)}(x)\Big|_{\alpha\leftrightarrow\beta}.\qquad
\end{eqnarray}

By means of Eq. (\ref{euler-1}) we find that 
\begin{eqnarray}\label{adendos-3-conv}
\begin{array}{l}
\mbox{the hypergeometric functions in }
\bar{{H}}^{(i)}  \; \text{ converge on }
|x|=1 \mbox{ if } \mbox{Re}\;\delta<1. \end{array}
\end{eqnarray}
%






%

%

%
%

%
%
%

\end{document}